\documentclass[11pt]{article}
\usepackage{amsmath,amssymb,amsthm,alg}

\makeatother
\newcommand{\ints}{\mathbb{Z}}

\newcommand{\ket}[1]{|#1\rangle}

\newcommand{\ignoretext}[1]{}

\newsavebox{\cardofbox}
\newlength{\cardofboxheight}
\newlength{\cardofboxdepth}
\newlength{\cardoftmpld}
\sbox{\cardofbox}{$l$}
\newlength{\cardoftmpls}
\sbox{\cardofbox}{$\scriptstyle l$}
\newlength{\cardoftmplss}
\sbox{\cardofbox}{$\scriptscriptstyle l$}
\newcommand{\concPow}[2]{#1^{\smallfrown #2}}

\newcommand{\seqIndex}[2]{#1_{#2}}
\newcommand{\cnot}[2]{\mbox{\textrm{c-not}}(#1,#2)}

\newcommand{\hadrot}[1]{\mathcal{H}(#1)}
\newcommand{\phase}[2]{\mathcal{R}_{#1}(#2)}

\newcommand{\assignfrom}{\leftarrow}

\newcommand{\qreverse}{\mbox{\textnormal{\textbf{reverse }}}}
\newcommand{\qifthen}[2]{\mbox{\textbf{\underline{if}~}}\;#1\;\textbf{\underline{then}~}\;#2}

\newcommand{\reversify}[2]{\subnameonly{#1}^{\textrm{R}}(#2)}
\newcommand{\assertClassical}[1]{\mbox{\textbf{isClassical~}}#1}

\newcommand{\dissipate}[2]{#1 \textbf{after dissipating~}#2}
\newcommand{\quantum}[1]{\underline{#1}}

\newcommand{\comment}[1]{\algbegin\mbox{\textbf{C: }}\textsl{#1}\\\algend}

\newcommand{\pproof}[1]{\algbegin\mbox{\textbf{Proof. }}\textsl{#1}\\\algend}
\newcommand{\subname}[2]{\mbox{\textsc{#1}}(#2)}
\newcommand{\subnameonly}[1]{\mbox{\textsc{#1}}}
\newenvironment{pseudocode}[1]{\par
\vspace{\baselineskip}
\setlength{\parindent}{0in}
\setlength{\parskip}{0in}
#1\par
}{
\vspace{\baselineskip}\par
}

\begin{document}

\title{Conventions for Quantum Pseudocode\\[.5cm]
{\large LANL report LAUR-96-2724}}
\author{E. Knill}
\date{June 1996}

\maketitle

\begin{abstract}
A few conventions for thinking about and writing quantum pseudocode
are proposed. The conventions can be used for presenting any quantum
algorithm down to the lowest level and are consistent with a quantum
random access machine (QRAM) model for quantum computing.
In principle a formal version of quantum pseudocode could be used
in a future extension of a conventional language.
\end{abstract}

\section{Introduction}
\label{section:introduction}

It is increasingly clear that practical quantum computing will take
place on a classical machine with access to quantum registers. The
classical machine performs off-line classical computations and
controls the evolution of the quantum registers by initializing them
in certain preparable states, operating on them with elementary
unitary operations and measuring them when needed.  Although
architectures for an integrated machine are far from established, a
suitable model for describing algorithms of this mixed nature
is that of the quantum random access machine (QRAM).  A quantum
random access machine is a random access machine in the traditional
sense with the ability to perform a restricted set of operations on
quantum registers. These operations consist of state preparation, some
unitary operations and measurement\footnote{
It may be convenient to allow application of general superoperators.
However any superoperator can be simulated by unitary operations and
measurement.
}. A QRAM can implement any local
quantum algorithm starting from classical states. Some situations
require operating on quantum registers in states prepared by another
source (for example a quantum channel, or a quantum transmission
overheard by an eavesdropper), in which case the QRAM is given
access to the required state in registers prepared elsewhere.

In classical computing, algorithms are often described using a loosely
defined convention for writing pseudocode. Good pseudocode is based on
computational primitives easily implemented in most computational
systems and has familiar semantics.  In principle, implementing
pseudocode on a real computer should require little effort. In
practice, pseudocode does not provide sufficiently detailed
implementations of the required data structures and often relies on
fairly high-level mathematical expressions, unbounded integers and
arbitrary accuracy real numbers. However, a good convention for
writing pseudocode is an indispensible tool for describing and
formally analyzing algorithms and data structures.

Conventions for writing quantum pseudocode have not yet been
established. In fact, most quantum algorithms are described using a
mixture of quantum circuits, classical algorithms and mathematical
description. Exceptions include algorithms
described in~\cite{beckman:qc1996a,cleve:qc1996a}.  The purpose
of this report is to provide some suggestions for unifying these
methods in a familiar framework.  The suggestions include methods for
handling quantum registers in pseudocode by introducing notation for
distinguishing quantum from classical registers, annotation for
specifying the extent of entanglement of quantum registers and methods
for initializing, using and measuring quantum registers.  On a higher
level, there are several meta-operations that have proved useful in
quantum computation. These include reversing a quantum operation not
involving a measurement, conditioning of quantum operations and
converting a classical algorithm to a reversible one. In this report
the meta-operations are described informally. Systematic
implementations of these algorithms will be given elsewhere.

The suggestions for writing quantum pseudocode are still incomplete.
The extent to which classical and quantum registers should be
separated and annotated remains to be seen. The most useful
meta-operations need to be better formalized in conjunction with a
more formal treatment of the syntax and the semantics of quantum
pseudocode. Conventions are likely to change as experience in writing
quantum algorithms is gained. Future versions of this report will
reflect such changes and include more examples.

\section{Quantum Pseudocode}
\label{section:qps}

\subsection{Quantum and Classical Registers}

Quantum pseudocode is an extension of conventional pseudocode
such as described in~\cite{cormen:qc1990a}.  The most important aspect
of the extension concerns the introduction and use of quantum
registers. We take the view that a quantum register is just a classical
register not known to be in a classical state.
The basic difference between a classical and a quantum
register is that the latter can be in a superposition of the available
classical states and allows only a restricted set of
operations. Except for the restriction on operations, the distinction
is primarily semantic.
If a register is known to be classical, all the usual operations familiar from
traditional programming can be applied to it. It is therefore convenient to
explicitly annotate those registers which may be in superpositions and
potentially entangled with other quantum registers.

The state of a machine executing quantum pseudocode can be described
by the contents of the classical registers and other classical
structures (such as program counters) required for the basic
architecture, together with a complex superposition of the classical
states of those registers that have been declared (explicitly or
implicitly) as quantum. An operation on any quantum register may have
an effect on the total superposition involving the other quantum
registers. Thus there are (with some exceptions) no side-effect free
operations on a quantum register\footnote{ An operation on a quantum
register will not affect the density matrix induced on the
others. However, the phases and the entanglement are modified, which
can affect the outcome of future operations.  }.  However, it may be
convenient to explicitly partition the quantum registers into sets
known to be in independent (that is factorizable) states. Two parts
must be merged whenever a unitary operation is applied involving both
of them.  They can be separated if it can be proven that the
operations result in a factorizable state under all circumstances.

\subsection{Methods for Introducing a Quantum Register}

The simplest method for introducing a quantum register is to do it
implicitly, by applying a proper unitary operation to a classical
register or by calling a subroutine which returns a quantum state.
Quantum registers can be distinguished by
underlines\footnote{Another option might be to use the notation $\ket{a}$
to indicate that register $a$ is participating in the quantum state of
the system. However, this is not quite consistent with the practice of
using $\ket{x}$ to denote the state labelled $x$.}.  The following
fragment of code gives examples:

\begin{pseudocode}{$\subname{QIntroExamples}{}$}
\begin{algtab*}
  $a\assignfrom\concPow{0}{5}$\\
  \comment{Initializes a classical register $a$ to contain $5$ bits
  in the $0$ state.}
  $\quantum{a}\assignfrom a$\\
  \comment{This converts $a$ to a quantum register without applying
  any operations. Future operations involving $\quantum{a}$ are
  restricted to quantum operations.}
  $d\assignfrom 10$\\
  \comment{$d$ is declared a classical register containing the
  integer $10$.}
  $\quantum{b}\assignfrom\algcall{UniformSuperposition}{d}$\\
  \comment{$\subname{UniformSuperposition}{d}$ takes a classical
  input (in this case an integer) and introduces a new quantum
  register in an initial state. Its state is independent
  of any other quantum register in the system.}
  $\quantum{c}\assignfrom\algcall{Multiply}{\quantum{b},d}$\\  
  \comment{Here a subroutine takes both quantum and classical input.
  A new quantum register $\quantum{c}$ is introduced, which
  may be entangled with $\quantum{b}$.}
  $x\assignfrom\algcall{DoSomethingClassical}{}$\\
  $\quantum{x}\assignfrom\algcall{DoSomethingQuantum}{x,d}$\\
  \comment{The fact that $x$ is converted and/or involved
  in quantum operation in the subroutine is made explicit
  by the assignment statement with the quantum annotation on the left.}
\end{algtab*}
\end{pseudocode}

The conventions used here require that a register symbol is always
considered either classical or quantum. Semantically, which is in
effect depends on the most recent operation applied to it. If it has
been declared as quantum, or a proper quantum operation has been
applied, then no further classical operations can be used until it is
measured.  The syntactic annotation helps keep track of the semantics
of a register in any given section of code. Correctness of the
annotation may require online consistency checking, though code where
the annotation is not obviously correct for syntactic reasons should
be avoided. Some of these issues can be avoided by using each register
in only one mode, either classical or quantum.  This convention is
used in~\cite{cleve:qc1996a}, where capitals are used to distinguish
quantum registers from classical ones. Note that if registers which
may be in a proper superposition are consistently annotated, then the
issue of whether classical registers are disjoint from quantum ones is
primarily a semantic issue.  In either case, annotation can be used to
indicate available knowledge on the nature of the superposition in a
register, e.g. whether the state of the register is purely classical
or not.

As can be seen from the code above, the idea is to use assignment with
quantum annotation on the left to indicate the introduction of quantum
registers or the conversion of classical registers to quantum
registers by subroutines or other operations. Thus assignments with a
quantum register on the left always have the property that the
register did not previously exist or is a classical argument on the
right. Further rules for assignments involving multiple types of
registers will be given when measurement is introduced. Although we
have not done so here, it may be desirable to distinguish such
generalized assignments from classical assignments by use of a different
left arrow.

\subsection{Applying Unitary Operators to Quantum Registers}

Operations on quantum registers are restricted to unitary operators
and measurement. Measurement is discussed in the next section.  Some
additional meta-operations will be given later.  Unitary operators can
only be applied to classical registers in a suitable conversion
statement, such as those introduced in the previous paragraphs. Which
unitary operators are applied can be controlled by the contents of
classical registers using the usual conditionals. Thus unitary
operations can be implicit in subroutine calls with both classical and
quantum arguments. What unitary operations are accessible must be
specified. In principle, any algorithm can be refined to one and two
qubit unitary operators~\cite{barenco:qc1995a}.  If the algorithm is
generic, any sufficiently powerful set may be chosen and the task of
reducing the algorithm to the device level can be left to the
implementer.  For efficiency purposes it might be worth specifying an
algorithm directly in terms of the most elementary operations
available in a given device, such as laser pulses for an ion trap
computer~\cite{cirac:qc1995a}.

If it is necessary to refine an algorithm to the qubit level, then it
is useful to have notation for extracting a (qu)bit from a register of
a given declared length. Indices are used
for this purpose.  As an illustration we give pseudocode of a
full quantum implementation of the Fourier transform mod $2^d$.

Define the Hadamard operation on one qubit as
\[
 \hadrot{\quantum{a}} =
\left(\begin{array}{cc}1&1\\1&-1\end{array}\right)\quantum{a}\,,
\]
The phase shift of $\ket{1}$ by $\phi$ is given by
\[
\phase{\phi}{\quantum{a}} = \left(\begin{array}{cc}1&0\\0&e^{i\phi}\end{array}\right)\quantum{a}\,.
\]
The controlled phase shift of $\quantum{b}$, controlled by $\ket{1}$
of $\quantum{a}$, is denoted by
\[
\qifthen{\quantum{a}}{\phase{\phi}{\quantum{b}}}\,.
\]
See below
for more information on quantum conditionals.

\begin{pseudocode}{\algname{Fourier}{$\quantum{a},d$}}
\alginout{
A quantum register $\quantum{a}$ with $d$ qubits. The most significant
qubit has index $d-1$.  }{ The amplitudes of $\quantum{a}$ are Fourier
transformed over $\ints_{2^d}$. The most significant bit in the output
has index $0$, that is the ordering is reversed.  }
\begin{algtab*}
  $\omega \assignfrom e^{i2\pi/2^{d}}$\\
  \algforto{$i=d-1$}{$i=0$}
     \algforto{$j=d-1$}{$j=i+1$}
       $\qifthen{\quantum{\seqIndex{a}{j}}}{\phase{\omega^{2^{d-i-1+j}}}{\quantum{\seqIndex{a}{i}}}}$\\
       \comment{
       If the phase change in this unitary operation is much smaller
       than then $1/n^2$, the operation can be omitted at a correspondingly
       small cost in the accuracy of the final
       state~\cite{coppersmith:qc1994a}. Thus this procedure can
       be modified to accept a precision parameter to reduce
       the number of quantum operations required.
       }
     \algend
     $\hadrot{\quantum{\seqIndex{a}{i}}}$\\
  \algend
\end{algtab*}
\end{pseudocode}

The pseudocode makes liberal use of
integer and real registers which are not explicitly represented
at the bit level. The data type of a register is implicit
in the first assignment statement which introduces it.

\subsection{Measuring a Quantum Register}

The most common method
for returning a quantum register to a classical state is to measure it.
The assignment statement $a\assignfrom\quantum{a}$ can
be used to indicate the measurement. The outcome of this operation
is inherently random and has side effects on the quantum state
of the part of the system previously entangled with $\quantum{a}$.
If a quantum input of a subroutine is measured in the subroutine
without being reintroduced as a quantum register, this can
be indicated by an assignment statement:
\[
a\assignfrom\algcall{DoAndMeasure}{\quantum{a},\quantum{b},c}\,.
\]
The most general assignment statement can have multiple registers
appearing on the left. The rules are as follows:
\begin{itemize}
\item[(i)]No register can appear in its quantum form on both sides.
\item[(ii)]A register appearing only
on the left must either be classical (in which case the original
contents are lost), or not previously declared.
\item[(iii)]A register appearing only on the right
can experience side effects during the operation.  That is registers,
in particular quantum registers, are assumed to be passed by reference.
It is a good idea to specify what side effects are experienced by
argument registers in the description of the output of the subroutine.
\item[(iv)]A register appearing in its quantum form on
the right and the classical form on the left is measured
during the operation, either explicitly, or implicitly at
the end.
\end{itemize}
For the purpose of clarity it is a good idea to indicate
all transitions between classical and quantum by use
of the generalized assignment statement.

As an example of the use of measurement for obtaining
a more efficient implementation, here is
pseudocode for the measured Fourier transform
described in~\cite{griffiths:qc1995a}.

\begin{pseudocode}{$a\assignfrom{\subname{MeasuredFourier}{\quantum{a},d}}$}
\alginout{
A quantum register $\quantum{a}$ with $d$ qubits. The most significant
qubit has index $d-1$.  }{ The amplitudes of $\quantum{a}$ are Fourier
transformed over $\ints_{2^d}$, and then measured.  The most
significant bit in the output has index $0$, that is the ordering is
reversed. The input quantum register is returned to a classical state
in the process.  }
\begin{algtab*}
  $\omega \assignfrom e^{i2\pi/2^{d}}$\\
  $\phi \assignfrom 0$\\
  \algforto{$i=d-1$}{$i=0$}
    $\phase{\phi}{\quantum{\seqIndex{a}{i}}}$\\
    $\hadrot{\quantum{\seqIndex{a}{i}}}$\\
    $\seqIndex{a}{i}\assignfrom\quantum{\seqIndex{a}{i}}$\\
    $\phi \assignfrom (\phi + \seqIndex{a}{i} \pi)/2$\\
    \comment{The expression on the right of this assignment
    statement requires $\seqIndex{a}{i}$ to be in a classical
    state as it involves operations not allowed for quantum registers.}
  \algend
\end{algtab*}
\end{pseudocode}

\subsection{Annotation of Quantum Registers}

As mentioned previously, it may sometimes be convenient
to explicitly annotate registers participating in independent
quantum states. There is no convention for this
at the moment. Suggestions include modifying the underline
which indicates a quantum register and various versions
of pre-, super-, and subscripts.

\subsection{Quantum Pseudocode without Annotation}

In principle, one can write quantum pseudocode without using
annotation. Note that only registers declared as bit sequences can be
used for quantum operations\footnote{ This may change as more
sophisticated quantum data structures are developed.}.  From an
operational point of view it suffices to describe what happens to a
register which is currently in superposition when subjected to a
classical (non-reversible) operation. The simplest solution is to
automatically force a measurement in that case and have the classical
operator act on the result. This includes assignment operations to
previously declared registers. An assignment operation involving a new
quantum register introduced by the subroutine replaces the target
register completely. Any quantum information is considered lost. The
operation is equivalent to a measurement with the outcome discarded
(\emph{dissipation} of the contents). An assignment operation
of a quantum to a classical register is a method for measuring
the quantum register and accessing the result elsewhere.

As can be seen, quantum pseudocode without annotation makes sense
operationally. However, correct annotation is helpful
for understanding and analyzing an algorithm. It also
helps the systematic application of some of the meta-operations
that are described next.

\subsection{Reversing a Quantum Subroutine}

One of the meta-operations frequently used in describing quantum
algorithms is reversal. This operation is usually introduced for
reversibly returning temporary registers to their starting state and
for reducing quantum register usage in converting a classical
algorithm to a reversible one (see
\cite{bennett:qc1989a,knill:qc1995b,beckman:qc1996a} for discussions
and examples).

A subroutine which performs quantum operations can be reversed
provided it does not contain measurement operations.
To perform the reversal, the classical
component of the subroutine must be implemented in a way
which explicitly keeps track of all unitary operators applied
to quantum registers. A simple method for doing this 
is to first run the subroutine forward without actually applying
any of the unitary operators, and then applying the inverses
of the unitary operators in reverse. In order to avoid
ambiguities, the forward subroutine must have no side effects
on classical input registers and no classical output.

The simplest case involves reversing a unitary operation.
For example, $\qreverse \hadrot{\quantum{a}}$
applies the inverse of the Hadamard transform to qubit $\quantum{a}$
(this happens to be the the same operation).
For a less trivial example, consider
the subroutine
\[
\quantum{b}\assignfrom\subname{Add}{\quantum{a},c}\,,
\]
which adds the contents of the classical register $c$ to $\quantum{a}$
and places the result coherently into $\quantum{b}$.
The operation
\[
\qreverse \quantum{b}\assignfrom\subname{Add}{\quantum{a},c}
\]
applies the inverses of quantum operations that would be
executed by the given subroutine in reverse order.
Thus the code

\begin{pseudocode}{$\subname{ReversingExample}{}$}
\begin{algtab*}
  $\quantum{a},\quantum{x},c\assignfrom\algcall{Initialize}{}$\\
  $\quantum{b}\assignfrom\subname{Add}{\quantum{a},c}$\\
  $\qifthen{\quantum{\seqIndex{b}{0}}}{\subname{DoStuffTo}{\quantum{x}}}$\\
  $\qreverse \quantum{b}\assignfrom\subname{Add}{\quantum{a},c}$
\end{algtab*}
\end{pseudocode}

returns $\quantum{b}$ to whatever classical state
it started in when it was introduced in the first call
to $\subname{Add}{}$. This does not normally hold if $\subname{Add}{}$
is replaced by an arbitrary quantum operation
or if the reverse of $\subname{Add}{}$
is applied to a general input state. In these cases $\quantum{b}$ may
end up in an entangled superposition.

\subsection{Quantum Registers in Provably Classical States}

In implementing quantum subroutines, it is often the case that
temporary quantum registers are introduced in such a way that at the
end (or at various other stages) they are known to have returned to a
classical state, at least if the operations are applied
perfectly. This often happens when transforming a classical algorithm
into a reversible form which is to be applied coherently to a quantum
input (see below).  It is
useful to be able to assert this fact (with proof if not obvious) and
explicitly return the register to the classical form without an actual
measurement.  (A measurement might make the subroutine apparently
non-reversible.)  The following (useless) fragment of code gives an
example.  Let $\cnot{a,b}$ be the controlled-not operation, controlled
by the first argument. Note that this is a classical reversible
operation and can therefore be used in both classical and quantum
contexts.

\begin{pseudocode}{\subname{IsClassicalExample}{}}
\begin{algtab*}
  $\quantum{a}\assignfrom\algcall{Generate}{}$\\
  $b\assignfrom 0$\\
  $\quantum{b}\assignfrom\cnot{\quantum{a},b}$\\
  $\quantum{c}\assignfrom\cnot{\quantum{b},c}$\\
  $\qreverse \quantum{b}\assignfrom\cnot{\quantum{a},\quantum{b}}$\\
    \comment{This is the same as $\quantum{b}\assignfrom\cnot{\quantum{a},\quantum{b}}$}
  $b\assignfrom\assertClassical{\quantum{b}}$\\
    \pproof{By checking on the classical states.}
\end{algtab*}
\end{pseudocode}

In cases where a register $\quantum{b}$ is provably in a classical
state, the contents are determined by the classical information
in the computation. Usually, as in the example above,
the contents are simply returned to a known initial state.
In either case, the register can be used again without
affecting reversibility of the code.

Another method for reusing a quantum register while maintaining our
ability to formally reverse a subroutine is to let its state
dissipate. This means that the contents of the register are no longer
needed, but also that any coherent information in it is lost with
possible side effects on the remaining quantum state.  To dissipate
and re-initalize $\quantum{b}$ one can use the statement
$\dissipate{b\assignfrom 0}{\quantum{b}}$.  Simply stating
$b\assignfrom 0$ in principle accomplishes the same thing (see
the section on annotation free pseudocode), but is not
explicit about the conversion of the quantum register. The important
property of such an operation is that the contents of the register
have no effect on future computations. The effect of reversing a
subroutine with such dissipation events but no measurements is still
predictable (the sequence of unitary operations applied can only
depend on the classical input, not on the quantum input), but not
equivalent to the inverse operation.  In effect
unitary operations involving the environment have occured and
these operations are not reversed when applying the inverse sequence
of unitary operators.  So far, allowing the contents of a register to
dissipate appears to be useful only if it is known to have returned to
a classical state already. Exceptions might be found in potential
applications to quantum non-deterministic
computing~\cite{knill:qc1996e}.

\subsection{Conditioning a Quantum Subroutine}

A frequently used technique, for example in reversible implementations
of classical functions, is to condition the application of a sequence
of unitary (or classical reversible) operations on the state of a
controlling qubit.  Provided that a subroutine is side effect free on
the classical input, has no classical output and avoids measurement
and dissipation, it can be conditioned by applying each of its unitary
operations if the controlling qubit is in $\ket{1}$ or applying the
identity if it is in $\ket{0}$. The overall effect is that of a
unitary operation involving the controlling qubit.  We can use a
version of the traditional if-then statement to perform quantum
conditioning. For example
\linebreak\begin{pseudocode}{}
$\qifthen{\quantum{b}}{\left(\begin{array}{cc}0&1\\1&0\end{array}\right)\quantum{a}}$\,,\\
$\qifthen{\quantum{c}}{\cnot{\quantum{b},\quantum{a}}}$
\end{pseudocode}
implement a controlled-not and a Toffoli gate, respectively.

Multiple conditionings can be efficiently implemented by
the use of an \emph{enable} qubit~\cite{beckman:qc1996a}.
An enable qubit is an auxiliary qubit which is introduced
specifically for controlling the quantum operations
in a subroutine.

\subsection{Making a Classical Function Reversible}

For most of the interesting quantum algorithms to date,
an important component is to entangle a quantum register
with the output of a function. This is
an operation of the form $\ket{a}\ket{0}\rightarrow\ket{a}\ket{f(a)}$,
where a classical algorithm for the function is known.
The classical algorithm cannot be applied directly because
it usually involves many non-reversible operations.
Techniques for systematically converting non-reversible
algorithms into reversible ones are known~\cite{bennett:qc1989a},
but usually do not yield the most space efficient implementation.
However, if complexity is not a major issue in describing an
algorithm, one can use a meta-operation on a classically
implemented function to indicate a reversible implementation.
If $b\assignfrom\subname{Function}{a}$ is given in a classical implementation
without side effects on $a$ or other classical inputs,
then $\quantum{b}\assignfrom\reversify{Function}{\quantum{a}}$
is interpreted as a reversible implementation with the
desired effect and all ancilla quantum registers reversibly
returned to their initial states.

\section{Acknowledgements}

This work was performed under the
auspices of the U.S. Department of Energy under Contract No. W-7405-ENG-36.

\bibliographystyle{plain}
\bibliography{journalDefs,qc}

\begin{thebibliography}{10}

\bibitem{barenco:qc1995a}
A.~Barenco, C.~H. Bennett, R.~Cleve, D.~P. DiVincenzo, N.~Margolus, P.~Shor,
  T.~Sleator, J.~Smolin, and H.~Weinfurter.
\newblock Elementary gates for quantum computation.
\newblock {\em Phys. Rev. A}, 52:3457--3467, 1995.

\bibitem{beckman:qc1996a}
D.~Beckman, A.~N. Chari, S.~Devabhaktuni, and J.~Preskill.
\newblock Efficient networks for quantum factoring.
\newblock {\em Phys. Rev. A}, 54:1034--, 1996.

\bibitem{bennett:qc1989a}
C.~H. Bennett.
\newblock Time/space trade-offs for reversible computation.
\newblock {\em SIAM J. Comput.}, 18:766--776, 1989.

\bibitem{cirac:qc1995a}
J.~Cirac and P.~Zoller.
\newblock Quantum computations with cold trapped ions.
\newblock {\em Phys. Rev. Lett.}, 74:4091--4094, 1995.

\bibitem{cleve:qc1996a}
R.~Cleve and D.~P. DiVincenzo.
\newblock Schumacher's quantum data compression as a quantum computation.
\newblock {\em Phys. Rev. A}, 54:2636--2650, 1996.

\bibitem{coppersmith:qc1994a}
D.~Coppersmith.
\newblock An approximate {F}ourier transform useful in quantum factoring.
\newblock Technical Report IBM Research Report 19642, IBM, 1994.

\bibitem{cormen:qc1990a}
T.~H. Cormen, C.~E. Leiserson, and R.~L. Rivest.
\newblock {\em Introduction to Algorithms}.
\newblock MIT Press, Cambridge, Mass, 1990.

\bibitem{griffiths:qc1995a}
R.~B. Griffiths and C-S Niu.
\newblock Semiclassical {F}ourier transform for quantum computation.
\newblock {\em Phys. Rev. Lett.}, 76:3228--3231, 1996.

\bibitem{knill:qc1995b}
E.~Knill.
\newblock An analysis of {B}ennett's pebble game.
\newblock Technical Report LAUR-95-2258, Los Alamos National Laboratory, 1995.
\newblock \texttt{http://lelimank.org/cv/reprints/knill:qc1995b.ps}.

\bibitem{knill:qc1996e}
E.~Knill.
\newblock Conventions for quantum pseudocode.
\newblock Technical Report LAUR-96-2724, Los Alamos National Laboratory, 1996.

\end{thebibliography}

\end{document}